\documentstyle[12pt,epsfig]{article}
\newcommand{\be}{\begin{equation}}
\newcommand{\ee}{\end{equation}}

\begin{document}

\title{Partial widths $a_0(980) \to \gamma\gamma$,
$f_0(980) \to \gamma\gamma$ and \\
$q \bar q $-classification of the lightest scalar mesons}
\author{A.V. Anisovich$^{(a)}$, V.V. Anisovich$^{(a)}$,
D.V. Bugg$^{(b)}$, and V.A. Nikonov$^{(a)}$ \\ \\
$^{(a)}$ St.Petersburg Nuclear Physics Institute,
Gatchina, 188350 Russia,\\
$^{(b)}$ Queen Mary and Westfield College,
Mile End Rd., London E1 4NS, UK}
\date{}
\maketitle

\begin{abstract}
We calculate  partial widths for the decays $a_0(980) \to
\gamma\gamma$ and  $f_0(980) \to \gamma\gamma$ under the  assumption
that $a_0(980)$ and  $f_0(980)$ are members of the basic $1^3P_0 q\bar
q$ nonet. The results are in a reasonable agreement with data thus
giving an argument for a $q \bar q$ origin of these mesons. We also
calculate the $\gamma\gamma$ partial widths for the other scalar
mesons, members of the $2^3P_0 q\bar q$ nonet.
\end{abstract}

\section{Introduction}

The determination of the lightest scalar $q \bar q$ nonet is a problem
of  principal importance both for quark systematics and the search for
exotic states. The key query here is an understanding of the origin of
$a_0(980)$ and  $f_0(980)$ mesons: a study of decays $a_0(980) \to
\gamma\gamma$ and $f_0(980) \to \gamma\gamma$ is an imperative step in
the analysis of the structure  of $a_0(980)$ and $f_0(980)$ (see, for
example, \cite{barnes} and references therein).

Here we perform the calculation of the  scalar meson transition form
factors $a_0(980) \to \gamma^*(q^2)\gamma$ and $f_0(980) \to
\gamma^*(q^2)\gamma$ in the region of small $q^2$;  these form
factors, in the limit $q^2 \to 0$, determine the partial widths
$a_0(980)\to \gamma\gamma $ and $f_0(980) \to \gamma\gamma$. Our
calculation is based on the spectral representation technique
developed in \cite{ffPR} for a study of the pseudoscalar meson
transitions $\pi^0 \to \gamma^*(q^2)\gamma$, $\eta \to
\gamma^*(q^2)\gamma$ and $\eta' \to \gamma^*(q^2)\gamma$.

In the region of moderately small $q^2$, where Strong-QCD works, the
transition form factor $q\bar q$-meson $\to \gamma^*(q^2)\gamma$ is
determined by the quark loop diagram of Fig. 1a which is a convolution
of the $q\bar q$-meson and photon wave functions, $\Psi_{q\bar q}
\otimes \Psi_{\gamma}$.  The calculation of the process of Fig. 1a is
performed in terms of the double spectral representation over $q\bar
q$ invariant masses squared, $s=(m^2+k^2_{\perp})/\left ( x(1-x)\right
)$  and $s'=(m^2+k'^2_{\perp})/\left ( x(1-x)\right )$ where
$k^2_{\perp}$, $k'^2_{\perp}$ and $x$ are the light-cone variables and
$m$ is the constituent quark mass. Following \cite{ffPR}, we represent
the photon wave function as a sum of two components which describe the
prompt production of the $q\bar q$ pair at large $s'$ (with a
point-like vertex for the transition $\gamma \to q\bar q $,
correspondingly) and the production in the low-$s'$ region where the
vertex $\gamma \to q\bar q $ has a nontrivial structure due to soft
$q\bar q$ interactions.  The process of Fig. 1a at moderately small
$|q^2|$ is mainly determined by the low-$s'$ region, in other words by
the soft component of the photon wave function.

The soft component of the photon wave function was restored in
\cite{ffPR} on the basis of the experimental data for the transition
$\pi^0 \to \gamma^*(q^2)\gamma$ at $|q^2| \leq 1$ GeV$^2$.  With the
photon wave function found, the form factors $a_0 \to
\gamma^*(q^2)\gamma$ and $f_0 \to \gamma^*(q^2)\gamma$ at $|q^2| \leq
1$ GeV$^2$ provide the opportunity to investigate in detail the scalar
meson wave functions.  However, the current data do not allow to
perform a full analysis, so we restrict ourselves by the consideration
of a one-parameter representation of the wave function of scalar
mesons, this parameter being  the mean radius squared $R^2$.

Within the assumption about $q\bar q$ structure of the lightest scalar
mesons, the flavour content of $a_0(980)$ is fixed, thus allowing
unambiguous calculation of the transition form factor $a_0(980) \to
\gamma\gamma$.  We obtain  reasonable agreement with data at
$R^2_{a_0(980)} \sim (11-27)$ GeV$^{-2}$  or, in terms of the pion
radius squared, at $R^2_{a_0(980)}/R^2_{\pi} \sim 1.1-2.7$.

The partial width $\Gamma (f_0(980) \to  \gamma\gamma)$ depends on the
relative weight of the strange and non-strange components of the
scalar/isoscalar meson, $s\bar s$ and $n\bar n$.  For the region of
not very large $R^2_{f_0(980)}$, $R^2_{f_0(980)}/R^2_{\pi} \sim
1.0-1.7$, the agreement with  data is attained at relatively large
$s\bar s $ component in $f_0(980)$, that is, of the order of $40-60\%
$. It does not contradict the results of the analysis of two-meson
spectra \cite{km1900} according to which the lightest
$(IJ^{PC}=00^{++})$-meson has a large $s\bar s$-component.

\section{Decay amplitude and partial width}

Below we present  the formulae for the scalar/isoscalar meson decay
$f_0 \to \gamma\gamma$. The formulae for $a_0 \to \gamma\gamma$
coincide in their principal points with those for
$f_0 \to \gamma\gamma$.

The amplitude for the scalar meson two-photon decay has the following
structure:
\be
A_{\mu\nu}= e^2
  g_{\mu\nu}^{\perp\perp} \; F_{f_0 \gamma\gamma}(0,0)\,.
\ee
Here $e$ is the electron charge ($e^2 /4\pi =\alpha = 1/137$)
and $F_{f_0 \gamma\gamma}(0,0)$
is the form factor for
the transition $f_0 \to \gamma(q^2)\gamma(q'^2)$
at  $q^2 =0$ and $q'^2 =0$, namely, $  F_{f_0 \gamma\gamma}(0,0) =
F_{f_0 \gamma\gamma}(q^2 \to 0,q'^2\to 0)$.

The metric tensor $ g_{\alpha\beta}^{\perp\perp}$  determines
the space perpendicular to $q$ and $q'$:
 \be
 g_{\alpha\beta}^{\perp\perp}=
 g_{\alpha\beta} -q_{\alpha}q_{\beta}\frac {q'^2}{D}
  -q'_{\alpha}q'_{\beta}\frac {q^2}{D}
  +(q_{\alpha}q'_{\beta}+q'_{\alpha}q_{\beta})\frac {(qq')}{D}\,,
\ee
where $  D=q^2q'^2-(qq')^2 $.

\subsection{Partial width}

The partial width, $\Gamma_{f_0 \to \gamma \gamma}$, is determined as
\be
m_{f_0}\Gamma_{f_0 \to \gamma \gamma} =
\frac 12  \int d\Phi_2(p_{f_0};q,q')
\Sigma_{\mu\nu} |
A_{\mu\nu} |^2 =
\pi \alpha^2
|F_{f_0 \gamma\gamma}(0,0)|^2\,.
\ee
Here $ m_{f_0}  $ is the $f_0$-mass,
the summation is carried over outgoing
photon polarizations, the photon identity factor, $\frac 12$,  is written
explicitly,  and the two-particle invariant phase space is equal to
\be
d\Phi_2(p_{f_0};q,q') = \frac 12 \; \frac {d^3q}{(2\pi)^3 2q_0}   \;
\frac {d^3q'}{(2\pi)^3 2q'_0}\;
(2\pi)^4\delta^{(4)} \left ( p_{f_0} -q-q' \right ).
\ee

\subsection{Form factor $F_{f_0 \gamma\gamma}(q^2,q'^2)$}

Following the prescription of Ref. \cite{ffPR},
we present the amplitude of the process of Fig. 1a in terms of the
spectral representation in the $f_0$ and
$\gamma (q')$ channels.
The double spectral representation reads
\be
F_{f_0 \gamma\gamma}(q^2 ,q'^2) =
 2\int \limits_{4m^2}^{\infty}
   \frac {ds\;ds'}{\pi^2}  \frac {G_{f_0}(s)}{s-m_{f_0}^2}
\ee
$$
\times  d\Phi_2(P;k_1,k_2) \;   d\Phi_1(P\; ';k'_1,k_2) \;  Z_{f_0}
   T(P\; ^2,P\; '^2,q^2) \sqrt {N_c}   \;
    \frac {G_{\gamma \to q\bar q}(s')}{s'-q'^2}\,.
$$
In the spectral integral (5),
the momenta of the intermediate states differ from those of
 the initial/final states. The corresponding momenta for
intermediate states are
re-denoted as shown in Fig. 1b:
\be
  q \to P\;-P', \qquad q' \to P', \qquad p_{f_0} \to P \; ,
\ee
$$
   P^2=s, \qquad P\;'^2=s', \qquad  ( P\;'-P)^2=q^2.
$$
It should be stressed that $ P\;'-P \ne q$.
The two-particle phase space
$ d\Phi_2(P;k_1,k_2)$ is determined by Eq. (4), while the
one-particle space factor is equal to
\be
d\Phi_1(P\; ';k'_1,k_2) = \frac 12 \;
\frac {d^3k'_1}{(2\pi)^3 2k'_{10}}
(2\pi)^4\delta^{(4)} \left (P\; ' -k'_1-k_2 \right )  .
\ee
The factor $Z_{f_0}$ is determined by the quark content of the $f_0$
meson: it is equal to
$Z_{n\bar n}=(e_u^2+e_d^2)/\sqrt 2 $ for the $n\bar n$ component,
and $Z_{s\bar s}=e_s^2 $ for the $s\bar s$ component.
The factor $\sqrt {N_c}$, where
$N_c=3$ is the number of colours, is related to the normalization of
the photon vertex made in Ref. \cite{ffPR}.
We have two diagrams: with quark lines drawn clockwise and
anticlockwise; the factor $2$ in front of the
right-hand side of Eq. (5) stands for this doubling.  The vertices
$G_{\gamma \to n\bar n}(s')$ and $G_{\gamma \to s\bar s}(s')$ were
found in Ref. \cite{ffPR}; the wave function
$G_{\gamma \to n\bar n}(s)/s$ is shown in Fig. 2.

We parametrize the $f_0$-meson wave function in the exponential
form:
\be
\Psi_{f_0}(s)=\frac {G_{f_0}(s)}{s-m_{f_0}^2} =
   Ce^{-bs},
\ee
where $C$ is normalization constant, and the parameter $b$ can
be related to the $f_0$-meson radius squared.

\subsection{Spin structure factor $T(P^2,P'^2,q^2)$}

For the amplitude of Fig. 1b with transverse polarized photons,
the spin structure factor is fixed by the quark loop trace:
\be
Tr [\gamma^{\perp\perp}_{\nu} (\hat k'_1+m)
\gamma^{\perp\perp}_{\mu} (\hat k_1+m) (\hat k_2-m)]
=   T(P^2,P'^2,q^2) \;   g_{\mu\nu}^{\perp\perp}\, .
\ee
Here $\gamma^{\perp\perp}_{\nu}$
and $\gamma^{\perp\perp}_{\mu}$ stand for photon vertices,
$ \gamma^{\perp\perp}_{\mu} =
g_{\mu\beta}^{\perp\perp}\gamma_{\beta}$,
while $g_{\mu\beta}^{\perp\perp}$ is determined by Eq. (2)
with the following substitution
$q \to P\;-P'$ and $ q' \to P'$.
Recall that the momenta $k'_1$, $k_1$
and $k_2$ in (9) are mass-on-shell.

One has
\be
    T(s,s',q^2) =    -2m \left [
    4m^2-s+s'+q^2-\frac {4ss'q^2}{2(s+s')q^2-(s-s')^2-q^4}
    \right ].
\ee

\subsection{Light cone variables}

The  formula (5) allows one to make easily the  transformation to
  the light cone variables using the boost along the z-axis.
    Let us use the frame
  in which the initial $f_0$-meson is moving along the
    z-axis with the momentum
  $p\to \infty$:
  \be
  P=(p+\frac {s}{2p}, 0, p), \qquad
  P\; '=(p+\frac {s'+q^2_{\perp}}{2p}, \vec q_{\perp}, p).
  \ee
Then the transition form factor $f_0 \to \gamma(q^2) \gamma $ reads:
\be
    F_{f_0 \gamma\gamma} (q^2,0)=
   \frac {2Z_{f_0} \sqrt {N_c} }{16\pi^3}
 \int \limits_{0}^{1}
   \frac {dx}{x(1-x)^2}  \int d^2k_{\perp} \Psi_{f_0} (s) \Psi_{\gamma}
(s')
    T(s,s',q^2),
\ee
where $ x=k_{2z}/p$ , $ \vec k_{\perp}=  \vec k_{2\perp}$, and
the $q\bar q$ invariant masses squared are
\be
       s=\frac{m^2+k^2_{\perp} }{x(1-x)}, \qquad
    s'=\frac{m^2+(\vec k_{\perp}-x \vec q_{\perp})^2 }{x(1-x)} .
\ee

\subsection{Meson charge form factor}

In order to relate the   wave function parameters $C$ and $b$ of
Eq.(8) to the $f_0$-meson radius squared, we calculate the
meson charge form factor shown diagrammatically in Fig. 1c.  The
amplitude has the following structure
$$ A_{\mu}=(p_{f_0\mu}+ p'_{f_0\mu})F_{f_0 }(q^2),$$
where the meson charge form factor $F_{f_0 }(q^2)$ is a convolution of
the $f_0$-meson wave functions $\Psi_{f_0}\otimes\Psi_{f_0}$:

\be
   F_{f_0 }(q^2 ) = \frac {1}{16\pi^3}
 \int \limits_{0}^{1}
   \frac {dx}{x(1-x)^2}  \int d^2k_{\perp} \Psi_{f_0} (s) \Psi_{f_0} (s')
    S(s,s',q^2).
\ee
$S(s,s',q^2)$ is determined by the quark
loop trace in the intermediate state:
\be
Tr [ (\hat k_1+m)
\gamma^{\perp}_{\mu} (\hat k'_1+m) (\hat k_2-m)] =
 [ P'_{\mu} +P_{\mu} -
\frac{ s'-s   }{q^2 }  (  P'_{\mu} -P_{\mu}
 ) ] \;S(s,s',q^2),
\ee
where
\be
\gamma^{\perp}_{\mu} =
g^{\perp}_{\mu\nu} \; \gamma_{\nu}\; , \qquad
g^{\perp}_{\mu\nu} = g_{\mu\nu}-
  ( P'_{\mu} -P_{\mu}) (P'_{\nu} -P_{\nu} ) /q^2 \; .
\ee
 One has
 \be
 S(s,s',q^2)=
\frac{q^2(s'+s-q^2)(s'+s-q^2-8m^2)}
{2(s+s')q^2-(s'-s)^2-q^4} + q^2.
\ee
The low-$q_{\perp}^2$  charge form factor,
\be
F_{f_0 }(q^2 ) \simeq 1-\frac{1}{6}R^2 q_{\perp}^2 ,
\ee
determines the $f_0$-meson wave function parameters,
$C$ and $b$.

\subsection{First radial excitation states, $2^3P_0 q\bar q$}

Equation (8) stands for the wave function of the basic state;
the wave function of the
first radial excitation can be written within an exponential
approximation as
\be
   \Psi_{f_0}^{(1)}(s)=C_1(D_1s -1)e^{-b_1s}\,.
\ee
The parameter $b_1$ can be related to the radius of the radial
excitation state, then the values $C_1$ and $D_1$ are
fixed by the normalization and orthogonality requirements,
$\left[\Psi_{f_0}^{(1)}\otimes\Psi_{f_0}^{(1)}\right]_{q^2=0}=1$ and
$\left[\Psi_{f_0}      \otimes\Psi_{f_0}^{(1)}\right]_{q^2=0}=0$.

\section{Results}

Using Eqs. (8), (12) and
(19), we calculate  $\gamma\gamma$ partial widths
of the $1^3P_0 q\bar q$ and $2^3P_0 q\bar q$ mesons.

\subsection{Partial widths $a_0(980)\to\gamma\gamma$ and
 $a_0(1450^{+90}_{-20})\to\gamma\gamma$}

The partial width $\Gamma (a_0(980)\to\gamma\gamma)$ is determined by
the same equation as for $f_0(980)$-decay, Eq. (12),
with the only substitution $Z_{f_0}\to Z_{a_0}=(e^2_u -e^2_d)/
\sqrt{2}=1/(3\sqrt{2}) $. The value $\Gamma (a_0(980)\to\gamma\gamma)$
is shown in Fig. 3 as a function of $R^2_{a_0(980)}$.
Experimental study of $\Gamma (a_0(980)\to\gamma\gamma)$
was carried out in Refs. \cite{a0f0,a0}, the
averaged value is: $\Gamma (\eta \pi)\cdot
\Gamma (\gamma\gamma)/\Gamma_{total}=0.24^{+0.08}_{-0.07}$ keV \cite{PDG}.
Using $\Gamma_{total} \simeq \Gamma (\eta \pi)+\Gamma (K \bar K)$,
we have
$\Gamma (a_0(980)\to\gamma\gamma)=0.30^{+0.11}_{-0.10}$ keV.
The calculated value of $\Gamma (a_0(980)\to\gamma\gamma)$
agrees with data at $R^2_{a_0(980)}= 19\pm 8$ GeV$^{-2}$: this
value looks quite reasonable for a meson of the
$1^3P_0 q\bar q$ multiplet.

If $a_0(980)$ is a member of the basic
$1^3P_0 q\bar q$ multiplet, the scalar/isoscalar meson
$a_0(1450^{+90}_{-20}) $ is the first radial excitation meson,
a member of the $2^3P_0 q\bar q$ multiplet.
Figure 3b demonstrates the values of partial widths
$\Gamma (a_0(1450^{+90}_{-20} )\to\gamma\gamma)$:
in the calculation, we use
$m_{a0(1450^{+90}_{-20})} = 1535$ MeV
following the results of the analysis \cite{hep}
and put $R_{a_0(1450)}/R_{a_0(980)} \simeq 1.22 $
assumimg that radii of the mesons of
the $2^3P_0 q\bar q$ multiplet are larger
than those for $1^3P_0 q\bar q$.

The transition form factor $F_{a_0(1450)
\gamma\gamma}$ is a convolution
of two wave functions, $\Psi_{a_0(1450)}\otimes
\Psi_{\gamma}$, one of them being the wave function of the first radial
excitation changes the sign, see Eq. (19). This fact results in a
relative suppression of the decay
$a_0(1450^{+90}_{-20})\to\gamma\gamma$:  $\Gamma (a_0(1450^{+90}_{-20}
)\to\gamma\gamma)/ \Gamma (a_0(980 )\to\gamma\gamma) \sim 1/10 $.

\subsection{Partial $\gamma\gamma$-widths for scalar/isoscalar mesons
of $1^3P_0 q\bar q$ and $2^3P_0 q\bar q$ multiplets}

In the analysis of $f_0 \to \gamma\gamma$ decays, one should take
into account that the scalar/isoscalar mesons in the mass range $1-2$
GeV are mixtures of not only $n\bar n$ and $s\bar s$ components but the
gluonium state as well.  Therefore, the transition form factor
of the $f_0$-meson reads:
\be
F_{f_0 \gamma\gamma} = \cos \alpha \;
(\cos \phi \; F_{f_0(n\bar n) \gamma\gamma} +\sin \phi \;
F_{f_0 (s\bar s) \gamma\gamma})
\ee
where $\sin^2\alpha$ is the probability of the gluonium component in
$f_0$-meson, and $\phi $ is mixing angle for
$n\bar n$ and $s\bar s$ components:
$\psi^{flavour}_{f_0} =\cos \phi \; n\bar n +\sin \phi \; s\bar s $.
According to the estimations
of Ref. \cite{ZPhys}, $\cos^2 \alpha \sim 0.7-0.9$ for the
$f_0$-mesons of
$1^3P_0 q\bar q$ and $2^3P_0 q\bar q$ multiplets.

Figure 3c demonstrates the partial $\gamma\gamma$-widths
for $f_0(980)$ calculated under assumptions that $f_0(980)$ is
either a pure $n\bar n$ state (solid curve) or pure $s\bar s$
(dashed curve).
Experimental analyses give $\Gamma (f_0(980)\to\gamma\gamma)=
0.42\pm 0.06 \pm 0.18$ keV \cite{a0f0} and
 $\Gamma (f_0(980)\to\gamma\gamma)=
0.63\pm 0.14 $ keV \cite{f0}; the averaged value reads:
 $\Gamma (f_0(980)\to\gamma\gamma)
=0.56\pm 0.11$ keV
\cite{PDG}. Our calculation shows that this value can be easily
understood if $f_0(980)$ has a significant $s\bar s$ component.
For example, for $R^2_{f_0(980)}= 11$ GeV$^{-2}$ and $\alpha =0$,
the data can be described
either with $\phi \simeq -35^o$ or
$\phi \simeq 63^o$.
 The existence of a significant $s\bar s$-component
in $f_0(980)$ agrees with the results of analysis  \cite{km1900}
for the two-meson spectra.

Figure 3d shows  the $f_0(1500)\to \gamma\gamma$ partial widths
calculated for pure $n\bar n$ and $s\bar s$ components within the
assumption that these $q\bar q$ components belong to $2^3P_0$
multiplet. One can see a strong suppression of the $\gamma\gamma$ decay
mode for both components, $n\bar n$ and $s\bar s$. The origin of this
suppression is the same as for the decay $a_0(1450^{+90}_{-20})
\to\gamma\gamma$: this is an approximate orthogonality of the photon
and $(2^3P_0q\bar q)$-meson wave functions in the coordinate/momentum
space.

\section{Conclusion}

Here we continue the investigation of the meson
two-photon decays started in Ref. \cite{ffPR}, where the partial widths
$\pi \to \gamma\gamma$,
$\eta \to \gamma\gamma$ and $\eta' \to \gamma\gamma$ were calculated.
In the present paper, we have calculated
partial widths
$a_0(980) \to \gamma\gamma$ and
$f_0(980) \to \gamma\gamma$ assuming that the mesons $a_0(980)$ and
 $f_0(980)$ are members of the basic $1^3P_0 q\bar q$ multiplet:
the results are in a reasonable agreement with the data.
This supports the idea of $q\bar q$ origin of the scalar mesons
$a_0(980)$ and $f_0(980)$ and gives the argument that the lightest
scalar nonet is located near $1000$ MeV (see discussion in Refs.
\cite{Kpi,rev} and in references therein).

A successful description of the data is due to two principal points
taken into account in the calculation: (i) the spin structure and
relativistic corrections are included into consideration in the
framework of the relativistic light-cone technique, (ii) the
subprocess $q\bar q \to \gamma\gamma$, which was found previously in
\cite{ffPR}, is used in the numerical analysis.

We are graceful to L.G. Dakhno, D.I. Melikhov and A.V. Sarantsev for
useful discussions. The paper is suppored by grants RFFI 96-02-17934
and INTAS-RFBR 95-0267.

\begin{figure}[h]
\vspace{3cm}
\centerline{\epsfig{file=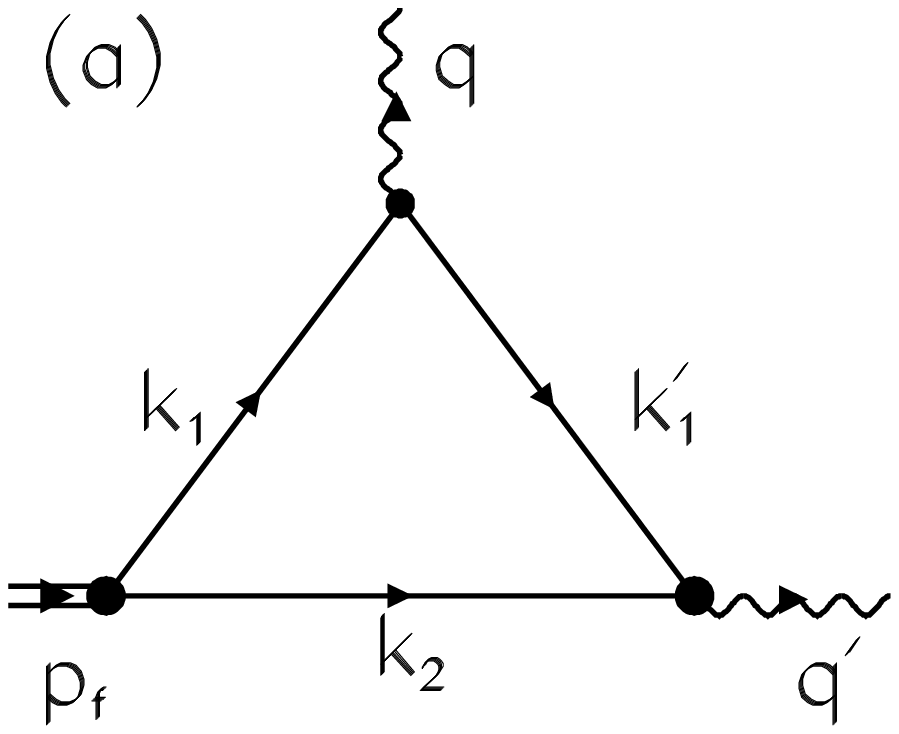,width=4.0cm}\hspace{1cm}
            \epsfig{file=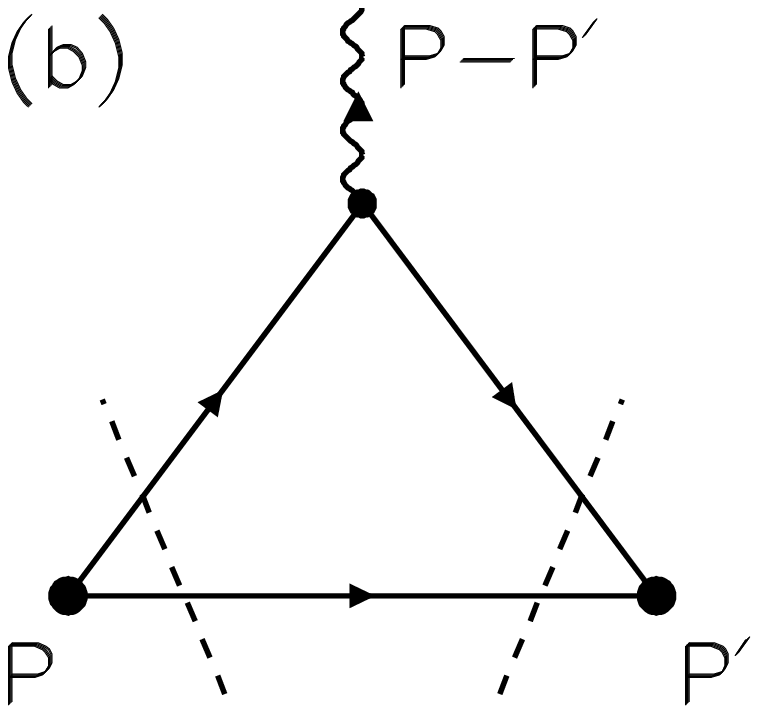,width=4.0cm}\hspace{1cm}
            \epsfig{file=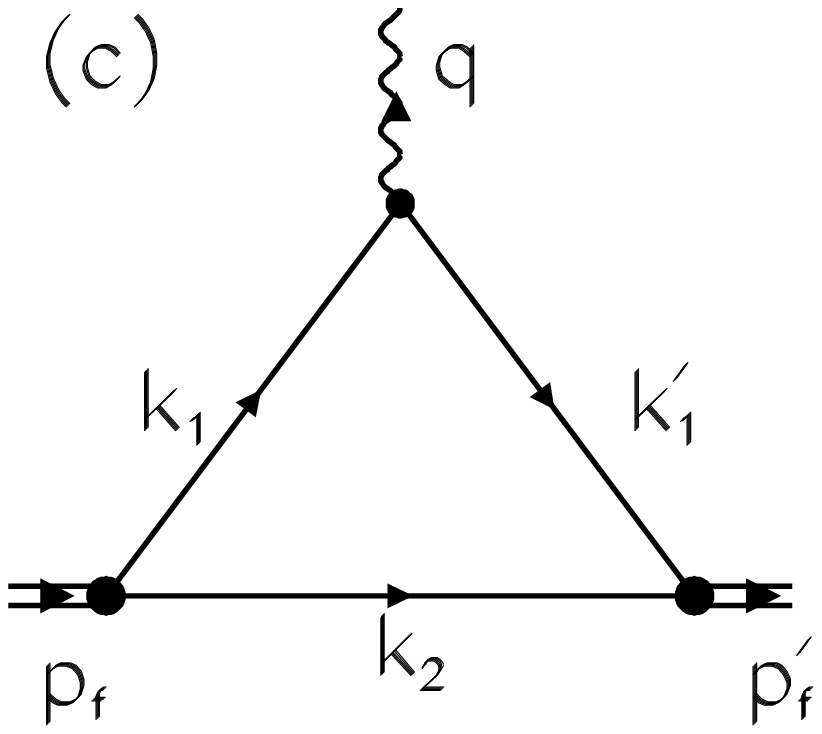,width=4.0cm}}
\caption{a) Triangle diagram for the transition form factor $f_0
\to \gamma (q^2) \gamma (q'^2)$. b) Diagram for the double spectral
representation over $P^2=s$ and $P'^2=s'$, Eq. (5); the intermediate
state particles are mass-on-shell, the cuttings of the diagram are
shown by dashed lines. c) Triangle diagram for the meson charge form
factor.}
\end{figure}

\begin{figure}
\centerline{\epsfig{file=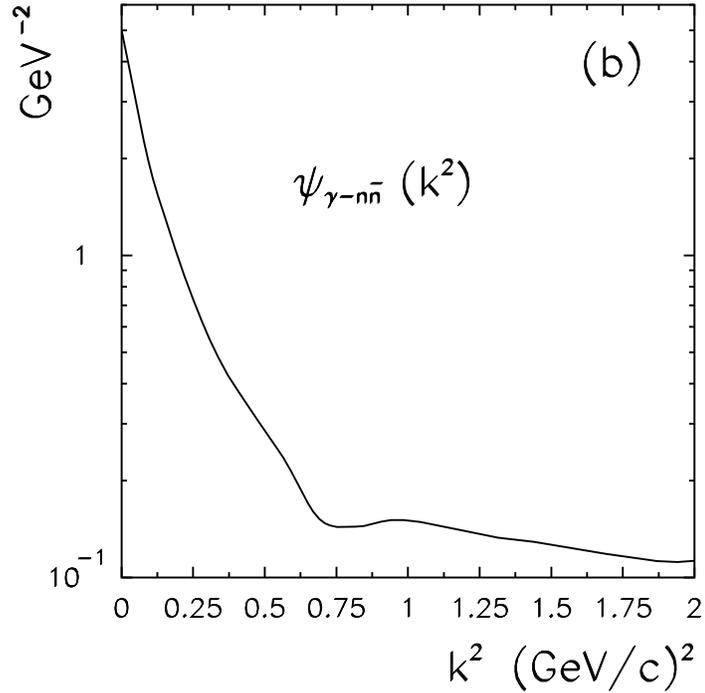,width=10cm}}
\caption{Photon wave function for non-strange quarks,
$\psi_{\gamma\to n\bar n}(k^2)=g_\gamma(k^2)/(k^2+m^2)$, where
$k^2=s/4-m^2$; the wave function for the $s\bar s$
component is equal to
$\psi_{\gamma\to s\bar s}(k^2)=g_\gamma(k^2)/(k^2+m^2_s)$ where $m_s$
is the constituent $s$-quark mass.}
\end{figure}

\begin{figure}
\centerline{\epsfig{file=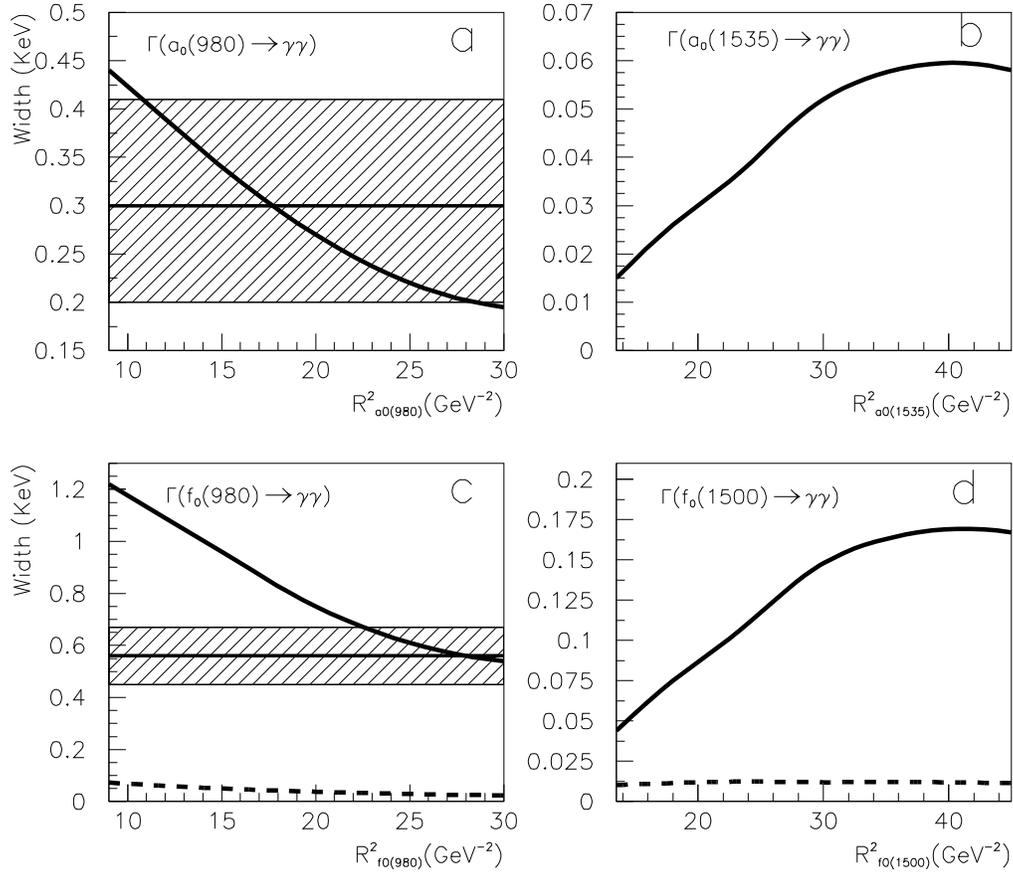,width=15cm}}
\caption{The calculated partial widths (solid and dashed curves)
versus meson radius squared and the experimental data
(scratched areas). For the scalar/isoscalar
mesons, $f_0(980)$ and $ f_0(1500)$, the dashed curves stand for the
pure $s\bar s$ content and the solid ones for the pure $n\bar n$. }

\end{figure}

\end{document}